\documentclass[aps,prl,twocolumn,showpacs,superscriptaddress]{revtex4}
\usepackage{graphicx}
\usepackage{amsmath}
\usepackage{amssymb}

\input{epsf}
\begin{document}

\title{Degeneracies in trapped two-component Fermi gases}

\author{K. M. Daily}
\affiliation{Department of Physics and Astronomy,
Washington State University,
  Pullman, Washington 99164-2814, USA}
\author{D. Rakshit}
\affiliation{Department of Physics and Astronomy,
Washington State University,
  Pullman, Washington 99164-2814, USA}
\author{D. Blume}
\affiliation{Department of Physics and Astronomy,
Washington State University,
  Pullman, Washington 99164-2814, USA}

\date{\today}

\begin{abstract}
We report on previously unobserved inter-system degeneracies
in two-component equal-mass 
Fermi gases
with interspecies zero-range interactions under isotropic
harmonic confinement.
Over the past 10 years,
two-component Fermi gases 
consisting of $n_1$ spin-up and $n_2$ spin-down atoms
with 
interspecies zero-range interactions have become a 
paradigm for modeling condensed matter systems, nuclear matter
and neutron matter.
We show that the eigen energies of the $(n_1+1,n_2-1)$
system 
are degenerate with the eigen energies of the $(n_1,n_2)$ system
for any $s$-wave scattering length $a_s$, including infinitely
large, positive and negative $a_s$.
The existence of the inter-system degeneracies is demonstrated explicitly for
few-body systems with $n_1+n_2=4, 5$ and 6.
The degeneracies and associated symmetries
are explained within a group theoretical framework.
\end{abstract}


\maketitle

Symmetry is one of the most fundamental concepts in physics,
underlying our understanding of elementary particle physics, relativity
and quantum mechanics, to name a few~\cite{zee}.
In quantum mechanics, symmetries manifest themselves in degeneracies
of energy eigen values.
If a Hamiltonian is invariant under rotations, for example, the 
eigen energies are $(2L+1)$-fold degenerate, where
$L$ denotes the orbital angular momentum quantum number~\cite{schiff}.
Similarly, the fact that the energy spectrum of the non-relativistic
hydrogen atom depends only on the principal quantum number
is intimately related to conserved quantities associated with 
the orbital angular momentum and Runge-Lenz vectors~\cite{schiff}.

Dilute atomic two-component Fermi gases with short-range interspecies
$s$-wave interactions can nowadays be realized routinely 
in many cold atom laboratories~\cite{kett08}.
% around the world
In these experiments, the atoms occupy two different hyperfine states
that are interpreted as spin-1/2 pseudo states.
Ultracold atomic Fermi gases have emerged as model systems
with which to study condensed matter phenomena such as
the BCS-BEC crossover 
and nuclear physics phenomena such as the 
equation of state of superfluid neutron 
matter~\cite{gior08,bloc08,geze11}.
%Over the past years, a
A multitude of results have been obtained for
two-component equal-mass Fermi gases with $s$-wave zero-range
(ZR) interactions.
A notable milestone is the derivation of various
universal relations by Tan~\cite{tan08a,tan08b,tan08c}, which are
centered around the ``contact'' and now form the basis
for novel spectroscopic techniques~\cite{stew10,wild11}.
Another notable milestone 
is the identification of a hidden SO(2,1) symmetry of the
two-component Fermi gas 
with ZR interactions at unitary in an isotropic harmonic trap
by Werner and Castin~\cite{wern06}, 
which manifests itself in ladders of uniformly
spaced excitation frequencies.

Our work 
identifies
another symmetry
that manifests itself in the existence of 
degenerate eigen energies of two-component 
equal-mass Fermi gases with the same number of particles
but different numbers of spin-up and spin-down atoms,
i.e., of $(n_1,n_2)$ and $(n_1',n_2')$ systems with 
$n_1+n_2=n_1'+n_2'$.
These ``inter-system degeneracies'' emerge in the ZR limit
for any value of the interspecies $s$-wave scattering length and
are broken for finite-range interactions or
unequal-mass systems.
%The degeneracies are explained within a group theoretical
%framework.

Our starting point is the non-relativistic 
Hamiltonian $H$ of the two-component Fermi gas 
with $n_1$ spin-up atoms and $n_2$ spin-down
atoms ($n=n_1+n_2$),
\begin{eqnarray}
\label{eq_ham}
H = H_0 + V_{\rm{int}},
\end{eqnarray}
where
\begin{eqnarray}
\label{eq_hamni}
H_0 = \sum_{j=1}^n \left( -\frac{\hbar^2}{2m} \nabla_{\vec{r}_j}^2 + 
\frac{1}{2} m \omega^2 \vec{r}_j^2 \right)
\end{eqnarray}
and $V_{\rm{int}}$ describes the interactions between the 
spin-up and spin-down atoms,
\begin{eqnarray}
\label{eq_vint}
V_{\rm{int}} = \sum_{j=1}^{n_1} \sum_{k=n_1+1}^{n} V_{\rm{tb}}(r_{jk}).
\end{eqnarray}
In Eq.~(\ref{eq_ham}), $m$ denotes the atom mass,
$\omega$ the angular trapping frequency,
and $\vec{r}_j$ the position vector of the $j$th atom
measured with respect to the trap center.
Following the literature~\cite{gior08},
the spin-up and spin-down components by themselves are 
assumed to be non-interacting (NI).
We model the intercomponent
atom-atom interactions 
by a short-range
Gaussian potential $V_{\rm{g}}$~\cite{stec07} 
with depth $V_0$ and range $r_0$,
$V_{\rm{g}}(r_{jk})=-V_0\exp[-r^2/(2 r_0^2)]$,
where $r_{jk}=| \vec{r}_j - \vec{r}_k|$.
For a fixed $r_0$, we adjust the depth $V_0$ such that
$V_{\rm{g}}$ reproduces the desired free-space zero-energy atom-atom
$s$-wave scattering length $a_s$.
We restrict ourselves to two-body potentials that
support no two-body $s$-wave bound state in free-space
for negative $a_s$ and one two-body $s$-wave bound state in free-space
for positive $a_s$.
In the $r_0 \rightarrow 0$ limit, our interaction model provides a
realization of the ZR $\delta$-function
interaction.
In practice, we determine the eigen energies of $H$ for a 
sequence of $r_0$ values and then extrapolate the eigen energies
to the $r_0 \rightarrow 0$ limit.
Throughout, we consider ranges $r_0$ that are much smaller than the
harmonic oscillator length $a_{\rm{ho}}$, 
where $a_{\rm{ho}}=\sqrt{\hbar/(m \omega)}$.

\begin{figure}
\vspace*{-0.8cm}
\includegraphics[angle=90,width=90mm]{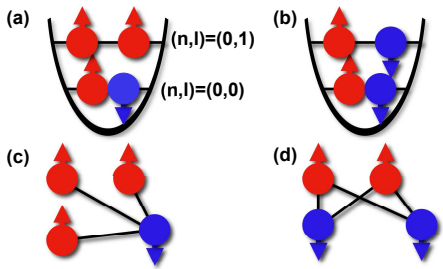}
\vspace*{-2.4cm}
\caption{(Color online)
Illustration of the $n=4$ systems.
Panels~(a) and (b) show the ground state configurations of the 
NI trapped
$(3,1)$ and $(2,2)$ systems. The horizontal solid
lines indicate the single particle harmonic oscillator orbitals with
energy $(2n+l+3/2)\hbar \omega$; the $(n,l)=(0,0)$ and $(0,1)$ orbitals
are respectively one-fold and three-fold degenerate.
Solid lines in panels~(c) and (d) 
illustrate the 
spin-up---spin-down
interactions of the $(3,1)$ and $(2,2)$ systems.
}\label{fig1}
\end{figure}

We first consider the Hamiltonian $H_0$, which describes
$n$ NI particles under isotropic harmonic confinement.
As an example, Figs.~\ref{fig1}(a) and \ref{fig1}(b) 
%schematically 
illustrate the ground state
configurations of 
the 
%$n=4$ systems with
$(n_1,n_2)=(3,1)$ and $(2,2)$ systems.
The lowest single particle orbital with 
energy $3 \hbar \omega/2$ can be occupied by
a spin-up atom and a spin-down atom.
To obey the Pauli exclusion principle, the other spin-up and
spin-down atoms need to occupy one of the three excited state
orbitals with energy $5 \hbar \omega/2$.
This simple picture yields a ground state energy of 
$8 \hbar \omega$ for both the $(3,1)$ and $(2,2)$ systems.
It can be readily shown that the ground state of
the $(3,1)$ system has $L^{\Pi}=1^+$ symmetry, where
$L$ denotes the orbital angular momentum quantum number
and $\Pi$ the parity; this ground state is three-fold degenerate
due to the rotational invariance of the Hamiltonian.
The ground state of the $(2,2)$ system is nine-fold 
degenerate~\cite{dail10}.
%;
%one state has $0^+$ symmetry, three states have 
%$1^+$ symmetry and five states have $2^+$ symmetry.
Just as the NI 
ground state manifolds of the $(3,1)$ and $(2,2)$ systems
contain degenerate energies corresponding to the 
same 
$L^{\Pi}$ symmetry,
so do the NI excited state manifolds.
Moreover, analogous degeneracies are readily identified for NI systems
with larger $n$.

In this paper, we are interested in the ``inter-system
degeneracies'', i.e., in the fact that the $(n_1+1,n_2-1)$ and 
$(n_1,n_2)$
systems support degenerate 
energies corresponding
to the same $L^{\Pi}$
symmetry.
Specifically,
we analyze what happens to the inter-system degeneracies
when the interactions are
turned on.
For example, 
since the $(3,1)$ system contains three 
spin-up---spin-down pairs while the 
$(2,2)$ system contains four 
[see Figs.~\ref{fig1}(c) and \ref{fig1}(d)],
it seems natural to expect that the interactions break the
inter-system degeneracies discussed above for the NI
$n=4$
systems. As we will show, however, this 
is not
the case if $r_0$ is taken to zero:
For ZR interactions, the eigen energies of the $(3,1)$ 
system form,
within our numerical accuracy, 
a subset of the eigen energies of the $(2,2)$
system.
Analogous results are found for systems with $n=5$ and $6$.

To determine the eigen energies of the Hamiltonian $H$
for finite depth $V_0$ of the Gaussian model
potential $V_{\rm{g}}$,
%---and thus
%for non-vanishing $s$-wave scattering
%length---, 
we resort to the stochastic variational
approach~\cite{stec07,cgbook}. 
We separate the center of mass motion and expand the relative 
eigen functions in terms of a basis set
with good orbital angular momentum
quantum number $L$ and parity $\Pi$~\cite{cgbook,suzu00,suzu08}.
The proper permutation symmetry under the exchange of identical
fermions is imposed by applying an appropriately
chosen anti-symmetrization
operator to the basis functions. 
Our implementation~\cite{raks12} allows for the treatment of
states with all $L^{\Pi}$ symmetries.
%vanishing and finite
%orbital angular momentum with either positive or negative parity.
%As discussed in Ref.~\cite{cgbook}, the
The
stochastic variational
approach results in variational upper bounds to the
exact 
eigen energies~\cite{cgbook}.
%The eigen energies 
%can be determined with controlled accuracy
%since the basis set extrapolation error can be estimated
%reliably.

\begin{figure}
\vspace*{+.50cm}
\includegraphics[angle=0,width=65mm]{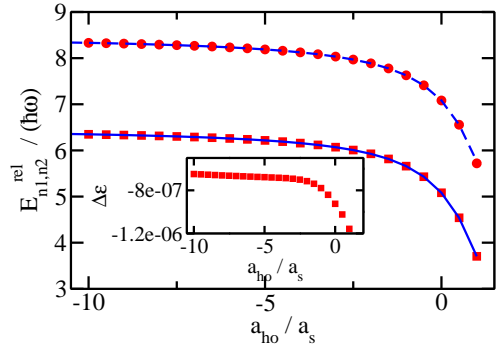}
\vspace*{-0.2cm}
\caption{(Color online)
Relative extrapolated ZR energies for the $n=4$ systems 
described by $H$ as a 
function of $a_{\rm{ho}}/a_s$.
Solid and dashed lines show the extrapolated ZR 
energies of the energetically lowest-lying and second lowest-lying
states of the $(3,1)$ system with $1^+$ symmetry
while squares and circles 
show 
those
of the $(2,2)$ system.
The energies of the $(3,1)$ and $(2,2)$ systems 
are indistinguishable on the
scale shown.
The inset shows the fractional difference $\Delta \epsilon$,
$\Delta \epsilon=(E^{\rm{rel}}_{2,2}-E_{3,1}^{\rm{rel}})/E_{3,1}^{\rm{rel}}$, 
for
the energetically lowest-lying
state.
}\label{fig2}
\end{figure}

As an example,
Fig.~\ref{fig2} shows the extrapolated ZR energies
for the $(3,1)$ and $(2,2)$ systems with $1^+$ symmetry
as a function of the inverse $s$-wave scattering length $a_s^{-1}$.
In this representation, the weakly-attractive BCS regime
($a_s <0$ and $|a_s|/a_{\rm{ho}} \ll 1$) is realized on the left of the
graph and the repulsive BEC regime ($a_s>0$)
on the right of the graph.
Lines show the relative eigen energies of the $(3,1)$ system and
symbols those of the $(2,2)$ system for $r_0=0$.
The extrapolated ZR energies are estimated to have
a combined basis set and ZR
extrapolation error smaller than $0.001$~\% 
for the energetically lowest-lying state
and smaller than 0.01~\% for the energetically
second lowest-lying state.
On the scale shown, the eigen energies of the $(3,1)$ and
$(2,2)$ systems are,
somewhat surprisingly, indistinguishable for both the lowest and 
second lowest states for all scattering lengths considered.
The inset shows that the fractional difference is 
smaller than $2 \times 10^{-4}$~\%
for the energetically lowest lying $1^+$ state
and the scattering lengths considered.
Thus, within our numerical accuracy, the energy curves agree 
throughout the crossover.

\begin{figure}
\vspace*{+.5cm}
\includegraphics[angle=0,width=65mm]{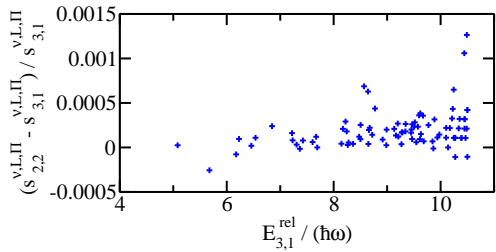}
\vspace*{-0.2cm}
\caption{(Color online)
Analysis of the extrapolated ZR energies of the $n=4$ 
systems described by $H$ at unitarity.
Symbols show the fractional difference 
$\Delta s$,
$\Delta s=(s_{2,2}^{\nu,L,\Pi}- s_{3,1}^{\nu,L,\Pi})/s_{3,1}^{\nu,L,\Pi}$,
between 
$s_{2,2}^{\nu,L,\Pi}$ and $s_{3,1}^{\nu,L,\Pi}$
as a function of $E_{3,1}^{\rm{rel}}$.
$\Delta s$ is of the order of or smaller
than the fractional numerical uncertainty of the extrapolated
ZR energies.
}\label{fig3}
\end{figure}

To see if the $(3,1)$ and $(2,2)$ energies are also degenerate
for other symmetries and for higher-lying excitations,
we focus on the infinite scattering length regime.
We analyze the extrapolated ZR energies of the $(3,1)$ and 
$(2,2)$ systems at unitarity for all states with relative
energy $E^{\rm{rel}}$ equal to or smaller than $21 \hbar \omega/2$, which
were determined in Ref.~\cite{raks12}
with an accuracy of 0.1~\% or better.
In this energy window, there exist 164 and 286 eigen energies
for the $(3,1)$ and $(2,2)$ systems, respectively~\cite{footnote0}. 
As pointed out by Werner and Castin~\cite{wern06}, the existence of a
hidden SO(2,1) symmetry leads to ladders of energies
spaced by $2\hbar \omega$, i.e., the relative eigen energies
at unitarity can be written as 
$E^{\rm{rel}} = (s^{\nu,L,\Pi} + 2q+1) \hbar \omega$, where
$q=0,1,\cdots$.
The separation constants $s^{\nu,L,\Pi}$ 
%values can be interpreted as 
%separation constants that 
arise when solving the $(n_1,n_2)$-fermion
problem within the hyperspherical framework.
We find that the relative eigen energies with 
$E^{\rm{rel}} \le 21 \hbar \omega/2$, corresponding to
$(3,1)$ and $(2,2)$ 
states that are affected by the interactions, are characterized
by 89 and 170 $s^{\nu,L,\Pi}$ values, 
respectively~\cite{footnote1}.
Quite surprisingly, every $s^{\nu,L,\Pi}$ value of the
$(3,1)$ system,
within
the numerical accuracy~\cite{raks12}, 
appears in the sequence of $s^{\nu,L,\Pi}$
values of the $(2,2)$ system.
Figure~\ref{fig3} shows that
the fractional difference between the
$s^{\nu,L,\Pi}$ values of the $(3,1)$ and $(2,2)$ systems 
is of the order of or smaller than the numerical
accuracy of the extrapolated ZR energies.
This suggests that the exact ZR energies of the 
$(3,1)$ system at unitarity form a subset of the exact
ZR energies of the $(2,2)$ system
at unitarity. These findings are corrobated by 
extensive perturbative calculations~\cite{supplement}.

\begin{figure}
\vspace*{+.5cm}
\includegraphics[angle=0,width=65mm]{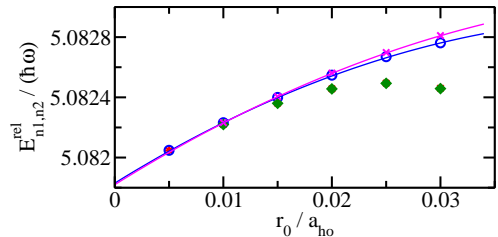}
\vspace*{-0.2cm}
\caption{(Color online)
Comparison of the eigen energies of 
the Hamiltonian $H$ and $H'$ 
with Gaussian model interactions as a function of $r_0/a_{\rm{ho}}$
at unitarity.
Crosses and circles show the relative eigen energies
$E_{n_1,n_2}^{\rm{rel}}$
of the energetically lowest-lying state
with $1^+$ symmetry
described by
$H$ 
while
diamonds and stars show 
the relative eigen energies
of the energetically lowest-lying state
with $1^+$ symmetry
described by
$H'$
(on the scale shown, the diamonds and stars are indistinguishable).
Solid lines show fits to the relative eigen energies of $H$.
}\label{fig4}
\end{figure}

The calculations presented so far strongly suggest that the
$(3,1)$ energies are degenerate with a subset of the $(2,2)$
energies in the $r_0 \rightarrow 0$ limit for all $a_s$.
The supplemental material~\cite{supplement} shows, 
using the stochastic variational and perturbative
approaches,
that analogous inter-system degeneracies exist for systems with $n=5$ and 
$6$.
To interpret our observations, we 
construct a new Hamiltonian $H'$,
\begin{eqnarray}
H'=H_0+V_{\rm{int}}',
\end{eqnarray}
that reproduces the eigen energies of the
$(n_1+1,n_2-1)$ and $(n_1,n_2)$ 
systems described by $H$ when $r_0 \rightarrow 0$.
The interaction potential $V_{\rm{int}}'$ includes 
interactions between all atom pairs and not just
between the spin-up and spin-down pairs,
\begin{eqnarray}
V_{\rm{int}}'=\sum_{j<k}^n V_{\rm{tb}}(r_{jk}).
\end{eqnarray}
The Hamiltonian $H'$ treats all atom pairs 
on equal footing. In particular, $V_{\rm{int}}'$ is the same 
for the $(n_1+1,n_2-1)$ and $(n_1,n_2)$ systems.
Intuitively, it is clear that the anti-symmetry of the 
eigen functions under the exchange of like atoms
``turns off'' the interactions between the like atoms when
$r_0 \rightarrow 0$, thereby ensuring that the energy spectra of $H$ and $H'$
are identical when $r_0 \rightarrow 0$.
This behavior is illustrated exemplarily in Fig.~\ref{fig4}
for the energetically 
lowest-lying state of the $n=4$ systems 
with $1^+$ symmetry 
interacting through
the Gaussian model potential.
In the $r_0 \rightarrow 0$ limit, the eigen energies of 
$H$ and $H'$ agree for the $(2,2)$ and $(3,1)$ systems;
moreover, as already pointed out above, the eigen energies
of the $(3,1)$ and $(2,2)$ systems agree.
For finite $r_0$, the 
eigen energies of the $(3,1)$ and $(2,2)$ systems
described by $H$ are characterized by different slopes
while the 
eigen energies of the $(3,1)$ and $(2,2)$ systems
described by $H'$ agree within our numerical accuracy
for all $r_0$.

The key motivation for introducing the Hamiltonian $H'$ is
that it describes all $n$-particle systems with ZR
$s$-wave interactions, regardless of the particle statistics.
The fact that the $(n_1+1,n_2-1)$ and $(n_1,n_2)$ systems are
described by the same Hamiltonian allows us to tie the 
evidenced degeneracy of the eigen energies to the
existence of a symmetry.
In particular, according to quantum mechanics~\cite{schiff}, 
the existence of degenerate eigen energies of a Hamiltonian is
a manifestation of an underlying symmetry. 
Since the Hamiltonian $H'$ is invariant under the permutation
of any pair of atoms, the inter-system degeneracies are intimately related to
the structure of the permutation group $S_n$.
Group theoretical tools are 
widely used in quantum chemistry and molecular physics 
to (anti-)symmetrize the wave functions
associated with the electronic and nuclear degrees of freedom~\cite{jensen}.
Here, they are employed to analyze the properties of the Hamiltonian $H'$, 
which has been shown to reproduce the eigen spectrum of 
the original
Hamiltonian $H$.

The Hilbert space of the $(n_1,n_2)$ system is spanned by
the direct product of the Hilbert spaces of the two single components
or, in terms of 
Young tableaux, 
$[1^{n_1}] \otimes [1^{n_2}]$~\cite{jensen,sakurai}.
Here, $[1^{n_1}]=[1,1,\cdots,1]$ indicates the fully anti-symmetric 
tableau of the
$n_1$ spin-up fermions.
The direct product can be decomposed into a 
direct sum of Young tableaux that
consist of at most two columns ($n_1 \ge n_2$)~\cite{explain1},
\begin{eqnarray}
\label{eq_young}
[1^{n_1}] \otimes [1^{n_2}]= 
[1^{n_1+n_2}] \oplus 
[2,1^{n_1+n_2-2}] \oplus 
\nonumber \\
\left[2^2,1^{n_1+n_2-4} \right] \oplus \cdots \oplus 
[2^{n_2},1^{n_1+n_2-2n_2}].
\end{eqnarray}
If we replace $n_1$ and $n_2$ in Eq.~(\ref{eq_young})
by $n_1+1$ and $n_2-1$, respectively, and then compare with
the decomposition for the $(n_1,n_2)$ state space,
we find that the decomposition of the $(n_1,n_2)$
state space contains the decomposition of the
$(n_1+1,n_2-1)$ state space,
\begin{eqnarray}
\label{eq_young2}
[1^{n_1}] \otimes [1^{n_2}]= 
\left(
[1^{n_1+1}] \otimes [1^{n_2-1}]
\right) \oplus
\left[2^{n_2},1^{n_1-n_2} \right].
\end{eqnarray}
This decomposition into irreducible representations
shows explicitly that the decomposition of the $(n_1+1,n_2-1)$
system is contained in that of the
$(n_1,n_2)$ system.
Correspondingly, the eigen energies of the $(n_1+1,n_2-1)$
system with ZR interactions
form a subset of those of the $(n_1,n_2)$ system
with ZR interactions 
for all $a_s$.
Equation~(\ref{eq_young2}) shows,
in agreement with our earlier discussion,
that
the $(n_1,n_2)$ system contains additional eigen energies.

In summary, we have identified and interpreted
inter-system degeneracies of two-component Fermi gases
with ZR interactions
under spherically symmetric confinement.
The fact that the eigen energies of the $n=4$ system with 
spin projection quantum number $M_S=1$ form a subset of the eigen energies
of the $n=4$ system with $M_S=0$ (and similarly for $n>4$) 
has multiple implications.
From a computational point of view, the degeneracies can be used to
test the accuracy of various schemes employed
to solve the $n$-fermion Schr\"odinger equation. 
Moreover, in certain cases it may be easier to treat the energetically
lowest lying state of a system with larger $M_S$ than an excited
state of a system with smaller $M_S$, allowing one to substitute
an
excited state calculation by a
ground state calculation for a system of the same size
but with different $M_S$.
The inter-system degeneracies also have experimentally observable
implications.
Since the change of the energy with scattering length
coincides for certain eigen states of the $(n_1-1,n_2+1)$
and $(n_1,n_2)$ systems, the corresponding
eigen states, which characterize two distinctly different
physical systems, have
the same 
contact~\cite{tan08a,tan08b,tan08c}.
Moreover, the two distinctly different systems share a
common set of eigen frequencies. These frequencies can
%, at least in principle,
be measured via microwave spectroscopy~\cite{selimjochim}.

The discussed inter-system degeneracies 
do not only exist
for systems with ZR interactions but also for systems with
finite-range interactions such as electronic systems, provided the
Hamiltonian under study is invariant under permutation
of all particle pairs.
This also implies that the degeneracies are not limited to
harmonically confined systems but also exist for systems in free 
space or under non-harmonic confinement, provided the
Hamiltonian under study is invariant under permutation. 
We conclude by noting that our analysis is based on the assumption
that the interaction potential is constructed from
pairwise two-body interactions.
The presence of three-body forces, 
which are needed to describe non-universal 
Efimov states or nuclear
systems, introduces a new degree of freedom not
considered here.

Support by the NSF through
grant PHY-0855332 
and by the ARO
is gratefully acknowledged.


\begin{thebibliography}{100}
\bibitem{footnote0}
%we note that t
These
numbers do not account for the $(2L+1)$-fold degeneracy of each 
eigen energy~\protect\cite{raks12}.
\bibitem{explain1}
Following standard notation, we note that $[2^2,1^0]=[2^2]=[2,2]$
while $[2^2,1^1]=[2,2,1]$.
\end{thebibliography}
\end{document}

\title{Support material for 
``Degeneracies in trapped two-component Fermi gases''}

\author{K. M. Daily}
\affiliation{Department of Physics and Astronomy,
Washington State University,
  Pullman, Washington 99164-2814, USA}
\author{D. Rakshit}
\affiliation{Department of Physics and Astronomy,
Washington State University,
  Pullman, Washington 99164-2814, USA}
\author{D. Blume}
\affiliation{Department of Physics and Astronomy,
Washington State University,
  Pullman, Washington 99164-2814, USA}

\maketitle

\date{\today}

This material summarizes details of our calculations and 
results that support the
conclusions drawn in the main part of the paper. 
Sections~\ref{sec_appendixa} and \ref{sec_appendixb}
show, within a perturbative
framework, that the inter-system degeneracies exist in the
weakly-interacting limits.
% with negative and positive 
%$s$-wave scattering length $a_s$.
Section~\ref{sec_appendixc} provides numerical examples for the inter-system
degeneracies for two-component gases 
at unitarity with $n=4$ and $5$ atoms.
Last, Sec.~\ref{sec_grouptheory}
discusses 
the application of 
the group theoretical framework discussed in the main text to
systems with up to $n=6$ atoms.

\section{Weakly-attractive Fermi gas}
\label{sec_appendixa}
When $|a_s|/a_{\rm{ho}} \ll 1$ and $a_s<0$,
the two-component Fermi gas can be described perturbatively assuming
$\delta$-function interactions between the spin-up and spin-down 
atoms. For the perturbative analysis
of this regime,
we assume that
the two-body potential $V_{\rm{tb}}(\vec{r}_{jk})$ in Eq.~(3)
of the main text is written as
$(4 \pi \hbar^2 a_s/m) \delta(\vec{r}_{jk})$.
Within first-order degenerate perturbation theory,
the relative energy $E^{\rm{rel}}_{n_1,n_2}$ of the $(n_1,n_2)$
system can be written in terms of
the non-interacting energy $E^{\rm{rel,ni}}_{n_1,n_2}$ and an energy
shift that is proportional to $a_s/a_{\rm{ho}}$~\cite{stec07b,stec08},
\begin{eqnarray}
\label{eq_energyshift}
E^{\rm{rel}}_{n_1,n_2} = 
E_{n_1,n_2}^{\rm{rel,ni}}+ 
(2 \pi)^{-1/2} c_{n_1,n_2} \frac{a_s}{a_{\rm{ho}}} \hbar \omega.
\end{eqnarray}
The coefficients $c_{n_1,n_2}$ are system and state dependent.
The non-interacting wave functions can be constructed
within the hyperspherical coordinate approach~\cite{dail10}
or by writing the non-interacting wave functions in terms of 
a product of determinants built using the 
single particle harmonic oscillator orbitals~\cite{stec07b,stec08}.
We checked explicitly for $n=4-6$ and all possible
$(n_1,n_2)$ combinations that the 
energy shifts are the same for the interaction potentials
$V_{\rm{int}}$ and $V_{\rm{int}}'$, supporting our finding that the
Hamiltonians $H$ and $H'$,
Eqs.~(1) and (4)
of the main text, are equivalent in the zero-range
limit.

Columns 2 and 3 of Table~\ref{tab_weak_n4}
show the coefficients $c_{2,2}$ and $c_{3,1}$
for the $(2,2)$ 
and $(3,1)$ systems with $L^{\Pi}=1^+$
symmetry and $E^{\rm{rel,ni}} \le 21 \hbar \omega/2$.
\begin{table}
\caption{Perturbative energy shifts for
the
$(2,2)$ and $(3,1)$ 
systems with $1^+$ symmetry in the weakly-attractive 
$a_s<0$ regime.
The coefficients $c_{2,2}$ and $c_{3,1}$ are
listed for the non-interacting energy manifolds 
$E^{\rm{rel,ni}}=13 \hbar \omega/2$, 
$17 \hbar \omega/2$ and $21 \hbar \omega/2$.
Within each energy manifold, the coefficients are listed
in decreasing order.
The $c_{3,1}$ column contains ``blank rows''
to highlight that only a subset of the
$(3,1)$ shifts appears in the
$(2,2)$ sequence.
The energy shifts for the $(2,2)$ system with 
$E^{\rm{rel,ni}} \le 17 \hbar \omega/2$ are taken from
Ref.~\cite{dail10}.
}
\begin{ruledtabular}
\begin{tabular}{ccc}
$E^{\rm{rel,ni}}/(\hbar \omega)$ & $c_{2,2}$ & $c_{3,1}$ \\
\hline
13/2 & $4$ & $4$ \\ \hline
17/2 & $4.50566$ & $4.50566$\\
 & $3$ & \\
 & $1.73167$ & $1.73167$ \\
 & $0.512668$ & $0.512668$ \\
 & $0$ & $0$ \\ \hline
21/2 & $4.96842$ & $4.96842$ \\
 & $4.75108$ & $4.75108$ \\
 & $4.30231$ & $4.30231$ \\
 & $4.09201$ & \\
 & $3$ & \\
 & $2.29699$ & $2.29699$ \\
 & $1.83376$ & $1.83376$ \\
 & $1.28299$ & \\
 & $1.10983$ & $1.10983$ \\
 & $0.534486$ & $0.534486$ \\
 & $0$ & $0$ \\
 & $0$ & $0$ \\
 & $0$ & $0$ \\
 & $0$ & $0$ \\
 & $0$ & $0$ \\
 & $0$ & $0$ \\
 & $0$ & \\
\end{tabular}
\end{ruledtabular}
\label{tab_weak_n4}
\end{table}
Table~\ref{tab_weak_n4}
shows that every energy shift of the $(3,1)$ system
corresponds to an energy shift of the $(2,2)$ system.
However, the $(2,2)$ system exhibits energy shifts that are not
present in the $(3,1)$ system.
We find similar results
for the $n=4$ systems with other $L^{\Pi}$ symmetries
and for the $n=5$ and $6$ systems.
As an example, Table~\ref{tab_weak_n6}
summarizes the energy shifts for the $n=6$ system
 for the first two
non-interacting energy manifolds with $E^{\rm{rel,ni}}=23 \hbar \omega/2$
and $25 \hbar \omega/2$.
\begin{table}
\caption{Perturbative energy shifts for
the $(3,3)$, $(4,2)$ and $(5,1)$ systems in the weakly-attractive 
$a_s<0$ regime.
The coefficients $c_{3,3}$, $c_{4,2}$ and $c_{5,1}$ are
listed for the two energetically lowest lying 
non-interacting energy manifolds. 
Within each energy manifold, the coefficients are listed
in decreasing order.
The $c_{4,2}$ and $c_{5,1}$ column contain ``blank rows''
to highlight that only a subset of the
$(3,3)$ shifts appears in the
$(4,2)$ sequence and that only a subset of the
$(4,2)$ shifts appears in the
$(5,1)$ sequence.
}
\begin{ruledtabular}
\begin{tabular}{ccccc}
$E^{\rm{rel,ni}}/(\hbar \omega)$ & $L^{\Pi}$ & $c_{3,3}$ & $c_{4,2}$ & $c_{5,1}$ \\
\hline
23/2 & $0^+$ & $11$ & & \\ 
     & $1^+$ & $8.5$ & $8.5$ & \\ 
     & $2^+$ & $9.5$ &  & \\
\hline
25/2 & $0^-$ & $7.75$ & $7.75$ &  \\
     & $0^-$ & $7$ & & \\
     & $0^-$ & $6.25$ & $6.25$ & $6.25$ \\
     & $1^-$ & $10.1049$ & &    \\
     & $1^-$ & $9.34613$ & $9.34613$ &    \\
     & $1^-$ & $8.58664$ & $8.58664$ &    \\
     & $1^-$ & $7.02013$ & &    \\
     & $1^-$ & $5.94223$ & $5.94223$ &    \\
     & $2^-$ & $9.125$ & &  \\
     & $2^-$ & $9.04636$ & $9.04636$ &  \\
     & $2^-$ & $8.5$ & &  \\
     & $2^-$ & $8.11599$ & $8.11599$ &  \\
     & $2^-$ & $6.71264$ & $6.71264$ &  \\
     & $2^-$ & $5.5$ & $5.5$ & $5.5$ \\
     & $3^-$ & $10.25$ & &  \\
     & $3^-$ & $8.59307$ & $8.59307$ &  \\
     & $3^-$ & $7.15693$ & $7.15693$ &  \\
     & $4^-$ & $8.25$ & &  \\
     & $4^-$ & $7.75$ & $7.75$ &  \\
\end{tabular}
\end{ruledtabular}
\label{tab_weak_n6}
\end{table}
Columns 3, 4 and 5 
of Table~\ref{tab_weak_n6} list the $c_{3,3}$, $c_{4,2}$ and
$c_{5,1}$ coefficients for all allowed $L^{\Pi}$ 
symmetries.
It can be seen that
every energy shift of the $(5,1)$ system
corresponds to an energy shift of the $(4,2)$ system.
Moreover, 
every energy shift of the $(4,2)$ system
corresponds to an energy shift of the $(3,3)$ system.
However, 
the $(3,3)$ system exhibits energy shifts that are not
present in the $(4,2)$ system,
and
the $(4,2)$ system exhibits energy shifts that are not
present in the $(5,1)$ system.

\section{Weakly-repulsive $a_s>0$ regime}
\label{sec_appendixb}
In the weakly-repulsive regime,
the two-component Fermi gas can be thought of as consisting 
of 
up to $n_2$ 
composite bosonic molecules and unpaired spin-up and spin-down
atoms. 
If no molecules are formed,
the perturbative treatment is the same as 
for the $a_s<0$ case discussed in the previous section.
When molecules form,
unpaired spin-up and spin-down fermions interact, as in the
$a_s<0$ case, through  the $\delta$-function
potential with $s$-wave atom-atom scattering length $a_s$
while spin-up atoms and molecules (and spin-down atoms and
molecules) interact through a $\delta$-function
potential with $s$-wave atom-dimer 
scattering length $a_{ad}$~\cite{stec07b,stec08,skor57,petr03};
in addition, the atom mass $m$ is replaced by $4m/3$
to account for the change of the mass of one of
the collision partners. 
If two or more molecules form, the molecule-molecule
interaction is modeled through the $\delta$-function
potential with $s$-wave dimer-dimer scattering length $a_{dd}$
and reduced mass $m$~\cite{stec07b,stec08,petr04}.
Using this effective interaction model, 
we have calculated the energy
shifts for the $n=4$, 5 and 6 systems
within first-order degenerate perturbation theory for 
good portions of the energy spectra.
We
find that the energy shifts of the 
$(n_1+1,n_2-1)$ system form a subset of the energy
shifts of the $(n_1,n_2)$ system for $n=4$, 5 and 6.
Moreover, the energy shifts of the 
$(n_1+2,n_2-2)$ system form a subset of the energy
shifts of the $(n_1+1,n_2-1)$ system for $n=6$.
In the following we provide a qualitative analysis of the
$a_s>0$ regime assuming at least one molecule forms.

\begin{figure}
\vspace*{+.9cm}
\includegraphics[angle=90,width=90mm]{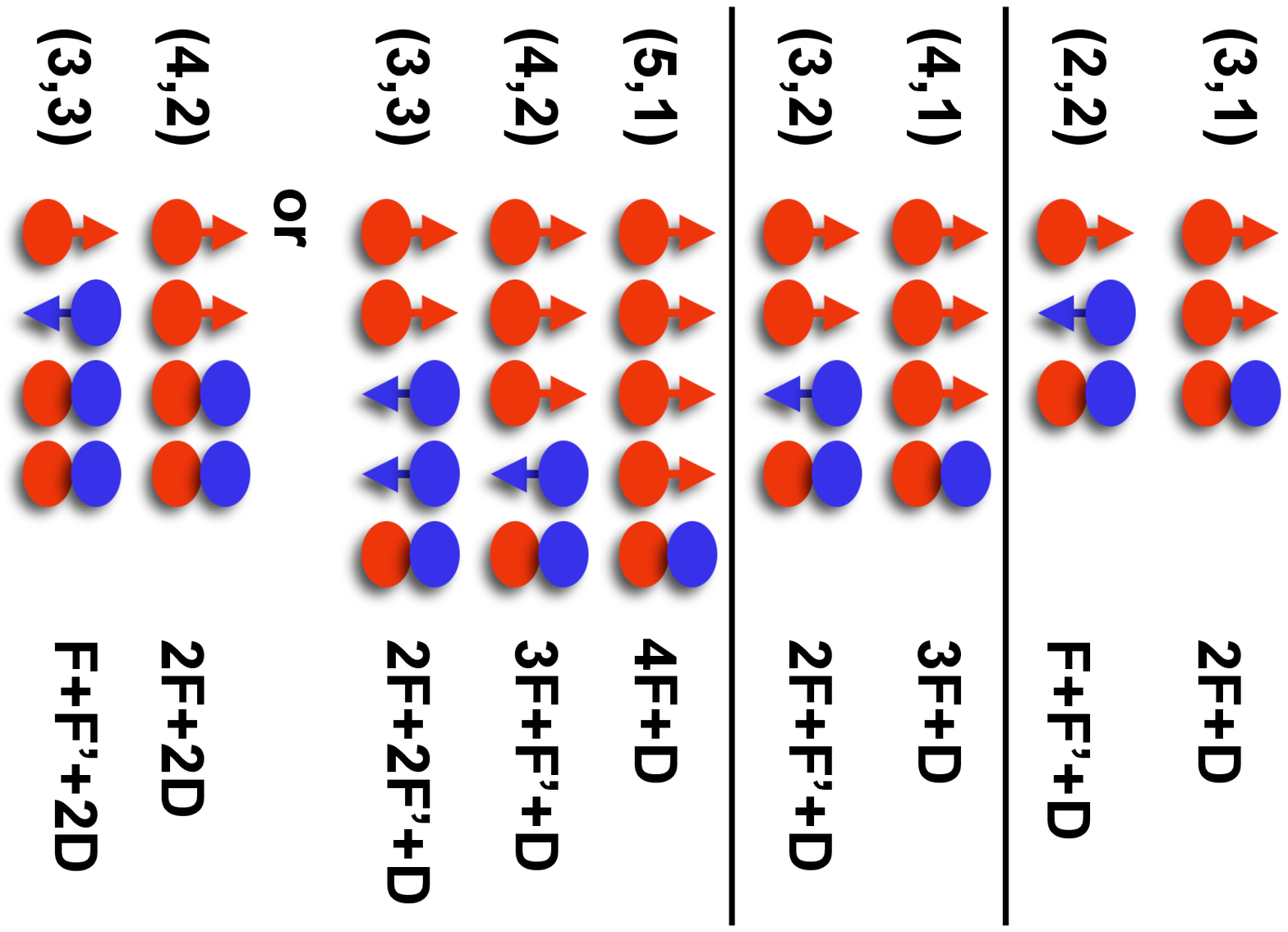}
\vspace*{0.2cm}
\caption{(Color online)
Schematic illustration of effective description
of two-component
Fermi gas in the weakly-repulsive
small $a_s/a_{\rm{ho}}$ regime ($a_s>0$).
Spin-up atoms (``F'') are depicted by red circles with 
up-pointing arrow.
Spin-down atoms (``F$'$'') are depicted by blue circles with 
down-pointing arrow.
Composite bosonic molecules or dimers (``D'') are depicted by
combined blue and red circles.
For the $(3,1)$, $(2,2)$, $(4,1)$, $(3,2)$ and $(5,1)$
systems, the formation of one molecule is considered.
For the $(4,2)$ and $(3,3)$ systems, the formation of one 
molecule
and two molecules is considered.
}\label{fig_effective}
\end{figure}

Figure~\ref{fig_effective}
schematically illustrates the effective systems
encountered for $n=4$, 5 and 6 when molecules are formed.
The $(3,1)$ system can form one molecule,
denoted by ``D'' (dimer)
while the $(2,2)$ system can form one or two
molecules. However, 
as we argue now,
only a portion of the energy spectrum
of the ``F+F$'$+D'' system (see Fig.~\ref{fig_effective})
is degenerate with the energy spectrum of the
``2F+D'' system.
While the wave function of the effective $(3,1)$ system
(the ``2F+D'' system)
must be anti-symmetric under the exchange of the two spin-up fermions,
no such constraint exists for the effective $(2,2)$ system
(the ``F+F$'$+D'' system).
This implies that the energy spectrum of the effective
$(3,1)$ system containing one 
molecule forms a subset of the energy
spectrum of the effective $(2,2)$
system containing one 
molecule. The analysis of the $(4,1)$ 
and $(3,2)$ systems proceeds similarly. In particular,
the ``3F+D'' system contains more symmetry constraints than
the ``2F+F$'$+D'' system. 
The number of atom-dimer interactions are, however,
the same. 
This implies that the energy spectrum of the effective
$(4,1)$ system containing one 
molecule forms a subset of the energy
spectrum of the effective $(3,2)$
system containing one 
molecule.

For the $n=6$ system, the situations where 
respectively one and two bosonic molecules form need
to be considered separately. 
We first consider the situation where one molecule
forms. The degeneracy of a part of the
energy spectrum of the
``3F+F$'$+D'' system with the energy
spectrum of the ``4F+D'' system can be understood along the same lines
as above. 
For the effective ``3F+F$'$+D'' and ``2F+2F$'$+D'' systems, 
in contrast, the situation is a bit more complicated. 
Intuitively, one may
argue that it
was
shown in Sec.~\ref{sec_appendixa} that the 
energy shifts of the ``3F+F$'$'' system form a subset of the
energy shifts of the ``2F+2F$'$'' system
and that the distinguishable ``D'' particle cannot ``destroy'' the 
degeneracy.
For the $(4,2)$ and $(3,3)$ systems, two molecules can form. Following the logic above, the effective ``2F+2D'' system
contains more constraints than the ``F+F$'$+2D'' system.
This implies that the energy spectrum of the effective
$(4,2)$ system containing two 
molecules forms a subset of the energy
spectrum of the effective $(3,3)$
system containing two 
molecules.

We reiterate that the qualitative pictures discussed here
have been confirmed by calculating the energy shifts for
good portions of the energy spectra
of the $n=4$, 5 and 6 systems.

\section{Unitarity}
\label{sec_appendixc}
Our numerical 
stochastic variational
approach for determining the eigen energies of small
trapped few-body systems under external spherical symmetric confinement,
interacting through finite-range Gaussian model potentials, 
is
discussed in detail in Ref.~\cite{raks12}.
For the discussion here it is important that this approach yields
the ground state and excited state 
energies of trapped few-body systems 
for any $s$-wave scattering length $a_s$ and that the 
accuracy of the eigen energies can be controlled systematically.

Reference~\cite{raks12} tabulates the extrapolated
zero-range energies of the 
$(3,1)$ and $(2,2)$
systems described by the Hamiltonian $H$, see Eq.~(1)
of the main text, at unitarity.
These energies and the associated $s_{n_1,n_2}^{\nu,L,\Pi}$ values were 
used to make Fig.~3
of the main text.
To exemplarily illustrate the degeneracy of
a subset of the $(2,2)$ energies and the $(3,1)$ energies,
Table~\ref{table_unit_n4}
summarizes the $(2,2)$ and $(3,1)$ energies 
\begin{table}
\caption{Extrapolated zero-range 
energies of the $(2,2)$ and $(3,1)$ systems
with $1^+$ symmetry at unitarity.
The $s^{\nu,L,\Pi}_{n_1,n_2}$ values are
related to the relative energies at unitarity through
$E_{n_1,n_2}^{\rm{rel}}=(2q + s^{\nu,L,\Pi}_{n_1,n_2}+1) \hbar \omega$,
where $q=0,1,\cdots$; the second and fourth columns list the
$s^{\nu,L,\Pi}_{n_1,n_2}$ values,
derived from the energies with
$q=0$. 
The energies are taken from Ref.~\cite{raks12}.
}
\begin{ruledtabular}
\begin{tabular}{cccc}
$E_{2,2}^{\rm{rel}}/(\hbar \omega)$ 
& $s_{2,2}^{\nu,L,\Pi}$ & $E_{3,1}^{\rm{rel}}/(\hbar \omega)$
& $s_{3,1}^{\nu,L,\Pi}$\\
\hline
5.0820 & 4.0820 & 5.0819 & 4.0819 \\
7.0822 &        & 7.0820 & \\
7.1895 & 6.1895 &        & \\
7.6060 & 6.6060 & 7.6056 & 6.6056 \\
8.1459 & 7.1459 & 8.1456 & 7.1456 \\
8.5000 & 7.5000 & 8.5000 & 7.5000 \\
8.9848 & 7.9848 & 8.9846 & 7.9846 \\
9.0827 &        & 9.0825 & \\
9.1341 & 8.1341 & 9.1324 & 8.1324 \\
9.1908 &        &        & \\
9.3750 & 8.3750 &        & \\
9.4561 & 8.4561 & 9.4544 & 8.4544 \\
9.6070 &        & 9.6060 & \\
9.6853 & 8.6853 & 9.6847 & 8.6847 \\
9.7798 & 8.7798 &        & \\
10.148 &        & 10.147 & \\
\end{tabular}
\end{ruledtabular}
\label{table_unit_n4}
\end{table}
with $E^{\rm{rel}} \le 21 \hbar \omega/2$
and $L^{\Pi}=1^+$
symmetry.
The second and fourth columns show the corresponding $s^{\nu,L,\Pi}$
values, derived from the energies with hyperradial quantum number $q=0$. 
It is easily checked that the relative energies 
of the $(3,1)$ system agree with the relative energies
of the  $(2,2)$ system to within the numerical uncertainty of 
0.1~\% of the
extrapolated zero-range energies.
Similarly good agreement is found for the
$n=4$ systems with other $L^{\Pi}$ symmetries.

Reference~\cite{raks12} also reports the
extrapolated zero-range
energies of the $(3,2)$ and $(4,1)$ systems at unitarity;
these energies have an estimated accuracy of 1~\%.
For example, the extrapolated
zero-range ground state energies of the 
$(4,1)$ system with $L^{\Pi}=1^+$, $2^+$ and $3^+$
symmetry described by the
Hamiltonian $H$ 
are
$7.155 \hbar \omega$,
$7.494 \hbar \omega$, and
$7.502 \hbar \omega$~\cite{raks12}.
For comparison, the extrapolated zero-range energy
of the second excited state of the
$(3,2)$ system with $L^{\Pi}=1^+$ symmetry is
$7.158 \hbar \omega$,
that of the fourth excited state of the
$(3,2)$ system with $L^{\Pi}=2^+$ symmetry is
$7.497\hbar \omega$,
and that of the second excited state of the
$(3,2)$ system with $L^{\Pi}=1^+$ symmetry is
$7.504\hbar \omega$.
These examples show that
the extrapolated zero-range energies of the $(4,1)$ system
described by $H$
form, within the numerical accuracy of Ref.~\cite{raks12}, 
a subset of the 
extrapolated zero-range energies
of the $(3,2)$ system described by $H$.

%For $n=6$...?????????? need to check my notes/calculations...
% listed in Ref.~\cite{raks12}
%support the notion that the findings for 
%the $n=4$ systems generalize to larger systems.
%Namely, the extrapolated zero-range energies
%of the $(4,1)$ system described by 
%the Hamiltonian $H$ form a subset of the extrapolated zero-range energies
%of the $(3,2)$
%system described by $H$.

\section{Application of group theoretical framework}
\label{sec_grouptheory}
%Th
%To explain the degeneracies of the 
%Hamiltonian $H'$, we employ
%zero-range
%interactions and
%treat the weakly-interacting regime
%perturbatively.
%The conclusions drawn from this analysis generalize
%beyond the perturbative regime
%and
%to finite-range interactions.

As discussed in the main text,
the Hamiltonian $H'$ is invariant under the exchange of any pair 
of particles and thus described by the symmetric group $S_n$
of degree $n$.
In the following, we apply Eqs.~(6) and (7) of the main text
to weakly-interacting atomic Fermi gases (small $|a_s|$) 
with $n=4, 5$ and 6
described by $H'$.

We first treat the $n=4$ system.
The $S_4$ group contains five conjugacy classes
or irreducible representations
that are characterized
by the Young tableaux $[4]$, $[3,1]$, $[2^2]=[2,2]$, $[2,1^2]=[2,1,1]$ and
$[1^4]=[1,1,1,1]$
with degrees $d_4=1$, $d_{31}=3$, $d_{22}=2$, $d_{211}=3$
and $d_{1111}=1$~\cite{sakurai}.
According to Eq.~(6) of the main text,
the properly anti-symmetrized states
of the $(2,2)$ and $(3,1)$
systems can be decomposed as
\begin{eqnarray}
\label{eq_young3}
[1,1] \otimes [1,1]= [1,1,1,1] \oplus [2,1,1] \oplus [2,2]
\end{eqnarray}
and
\begin{eqnarray}
\label{eq_young4}
[1,1,1] \otimes [1] = [1,1,1,1] \oplus [2,1,1],
\end{eqnarray}
respectively.
By inspection, this yields Eq.~(7) with $n_1=n_2=2$
of the main text,
\begin{eqnarray}
\label{eq_young5}
[1,1] \otimes [1,1] = ([1,1,1] \otimes [1]) \oplus [2,2].
\end{eqnarray}

To see how these decompositions relate to the energy spectra of the $(2,2)$
and $(3,1)$ systems, we determine the eigen energies of $H'$
for small $|a_s|/a_{\rm{ho}}$ perturbatively.
We start our discussion by considering 
the system consisting of {\em{four distinguishable}}
particles with equal masses, referred to as the XYZA system.
All six two-body interactions are assumed
to be characterized by the same $s$-wave scattering length.
Table~\ref{tab_grouptheory} lists the coefficients
$c$ [see Eq.~(\ref{eq_energyshift})], 
which determine the perturbative energy
shifts, for each $L^{\Pi}$ symmetry. 
\begin{table}
\caption{Perturbative analysis of the trapped $n=4$ system
with various symmetries
described by the Hamiltonian $H'$.
The first column lists the relative energy of the non-interacting 
energy manifold considered. The perturbative energy shifts
for a given angular momentum $L$
and parity $\Pi$ are reported in column three.
The energy shifts are calculated without imposing a specific 
symmetry on the wave functions; the corresponding degeneracies 
are listed in column four
(only states with projection quantum number
$M_L=0$ are considered). Columns five through 13 list the number
of linearly independent 
eigen states with a particular particle symmetry that can be constructed
from the unsymmetrized states that describe the XYZA system.
}
\begin{landscape}
\begin{tabular}{ c | c | c | c | c  c  c  c  c  c  c  c  c } 
$E^{\rm{rel,ni}}/(\hbar \omega)$ & $L^{\pi}$ & $c$ & XYZA & BBBB & BBBX & BBB$'$B$'$ & BBXY & BBFF & FFXY & FFF$'$F$'$ & FFFX & FFFF \\ 
\hline
$9/2$    & $0^+$ & 12      & 1& 1& 1& 1& 1&  &  &  &  &   \\
\hline
$11/2$   & $1^-$ & 8       & 3&  & 1& 1& 2& 1& 1&  &  &   \\
\hline
$13/2$ & $0^+$ & 6.5     & 2&  &  & 1& 1&  & 1& 1&  &   \\
       &       & 9       & 3&  & 1& 1& 2& 1& 1&  &  &   \\
       &       & 14      & 1& 1& 1& 1& 1&  &  &  &  &   \\\cline{2-13}
       & $1^+$ & 4       & 3&  &  &  & 1& 1& 2& 1& 1&   \\\cline{2-13}
       & $2^+$ & 5       & 2&  &  & 1& 1&  & 1& 1&  &   \\
       &       & 6       & 3&  & 1& 1& 2& 1& 1&  &  &   \\
       &       & 8       & 1& 1& 1& 1& 1&  &  &  &  &   \\
\hline
$15/2$ & $0^-$ & 0       & 1&  &  &  &  &  & 1& 1& 1& 1 \\\cline{2-13}
       & $1^-$ & 2.82461 & 3&  &  &  & 1& 1& 2& 1& 1&   \\
       &       & 4.42539 & 3&  &  &  & 1& 1& 2& 1& 1&   \\
       &       & 5       & 2&  &  & 1& 1&  & 1& 1&  &   \\
       &       & 5.12271 & 3&  & 1& 1& 2& 1& 1&  &  &   \\
       &       & 7.42288 & 3&  & 1& 1& 2& 1& 1&  &  &   \\
       &       & 9.20442 & 3&  & 1& 1& 2& 1& 1&  &  &   \\
       &       & 11      & 1& 1& 1& 1& 1&  &  &  &  &   \\\cline{2-13}
       & $2^-$ & 2       & 3&  &  &  & 1& 1& 2& 1& 1&   \\
       &       & 3       & 2&  &  & 1& 1&  & 1& 1&  &   \\
       &       & 4       & 3&  & 1& 1& 2& 1& 1&  &  &   \\\cline{2-13}
       & $3^-$ & 3.5     & 3&  &  &  & 1& 1& 2& 1& 1&   \\
       &       & 4       & 3&  & 1& 1& 2& 1& 1&  &  &   \\
       &       & 6       & 1& 1& 1& 1& 1&  &  &  &  &   \\
       &       & 6.5     & 3&  & 1& 1& 2& 1& 1&  &  &   \\
\end{tabular}
\end{landscape}
\label{tab_grouptheory}
\end{table}
Column 4 lists the number of
linearly independent eigen states with zero projection
quantum number, i.e., the number of
eigen states of the XYZA system with a given $L$, $\Pi$ and 
energy shift (and $M_L=0$).

To construct
eigen functions of four-particle systems with
a given permutation symmetry,
we perform a ``postsymmetrization'' of the
linearly independent eigen functions of the XYZA 
system.
There exist nine physical
four-body systems with distinct exchange
symmetry: BBBB, BBBX, BBB$'$B$'$, BBXY, BBFF, FFXY, FFF$'$F$'$
[referred to earlier as the (2,2) system], 
FFFX
[referred to earlier as the (3,1) system], 
and FFFF.
Here, ``B'' stands for ``boson'' and ``F'', as before, for ``fermion'';
the particle order does not matter (e.g., the XBBB is equivalent
to the BBBX system).
The number of properly symmetrized/anti-symmetrized
eigen states for these physical systems 
are listed in columns 5-13 of Table~\ref{tab_grouptheory}
for each
$L$, $\Pi$ and energy shift ($M_L=0$).
The number of linearly independent 
eigen functions with a given $L$, $\Pi$ and energy shift 
(and $M_L=0$) (see column four of Table~\ref{tab_grouptheory})
corresponds to the degree of the corresponding
irreducible representation of 
the $S_4$ group.

For the $9\hbar \omega/2$ energy manifold, e.g.,
one eigen state
exists. The ``highest symmetry'' of this eigen state corresponds
to the BBBB system, which is associated with the fully
symmetric Young tableau $[4]$. 
The BBBB state ``automatically'' also has the
proper symmetry of the BBBX, BBB$'$B$'$ and BBXY systems. 
The three linearly independent 
states in the $11 \hbar \omega/2$ energy manifold 
are associated with the Young tableau $[3,1]$.
Appropriate linear combinations of these linearly
independent functions give the properly symmetrized/anti-symmetrized
eigen functions of the 
BBBX, BBB$'$B$'$, BBXY, BBFF and FFXY
systems.
The two linearly independent eigen states
in the $13 \hbar \omega/2$ energy manifold with $L^{\Pi}=0^+$
and $c=13/2$ 
are associated with the Young tableau $[2,2]$.
Appropriate linear combinations of these linearly
independent functions give the properly symmetrized/anti-symmetrized
eigen functions of the 
BBB$'$B$'$, BBXY, FFXY and FFF$'$F$'$
systems.
Lastly, the 
three linearly independent 
states in the $13 \hbar \omega/2$ energy manifold 
with $L^{\Pi}=1^+$
are associated with the Young tableau $[2,1,1]$.
Appropriate linear combinations of these linearly
independent functions give the properly symmetrized/anti-symmetrized
eigen functions of the 
BBXY, BBFF, FFXY, FFF$'$F$'$ and FFFX
systems.

The perturbative calculations summarized above show that the
eigen states of the FFFX system are, as predicted by
Eq.~(\ref{eq_young4}), associated with either
the $[1,1,1,1]$ or the $[2,1,1]$ 
Young tableaux. In both cases, there exists a corresponding
eigen state of the FFF$'$F$'$ system
[see also Eq.~(\ref{eq_young3})]. It follows that
the eigen energies of the FFFX and FFF$'$F$'$ systems are
degenerate.
However, the perturbative calculations
show---in
agreement with Eqs.~(\ref{eq_young3})
and (\ref{eq_young5})---that there also exist eigen states of 
the FFF$'$F$'$ system that are associated
with the $[2,2]$ Young tableau. 
No corresponding eigen states 
with FFFX symmetry exist in this case, explaining the 
fact that the FFF$'$F$'$ system supports a set of eigen energies
that are absent in the FFFX spectrum. 
In selected cases, we have checked that the post-symmetrization
approach,
applied to the unsymmetrized
eigenstates of the XYZA system
at unitarity,
reproduces the energies of the $(2,2)$ and $(3,1)$ systems
reported in Ref.~\cite{raks12}.
Similarly, 
we tied the energy spectra of the
$(4,1)$, $(3,2)$, $(5,1)$, $(4,2)$ and $(3,3)$ systems
obtained perturbatively and by the stochastic variational
approach to
the irreducible representations of the symmetric groups 
$S_5$ and $S_6$.

\section{Acknowledgement}
Support by the NSF through
grant PHY-0855332 
and by the ARO
is gratefully acknowledged.